\documentclass[10pt, letterpaper]{article}

\usepackage{sagathm}
\usepackage{sagageneralarxiv}
\usepackage{float}
\usepackage[skip=2pt,font=small]{caption}

\renewcommand{\vec}[1]{\mathbf{#1}}

\newcommand{\nzd}{\operatorname{nzd}}
\newcommand{\Verts}{\scr{V}}
\newcommand{\Tris}{\scr{T}}
\newcommand{\Regs}{\scr{R}}
\newcommand{\Cost}{\scr{C}}
\newcommand{\Err}{\scr{E}}
\newcommand{\Dist}{\scr{D}}
\newcommand{\werr}{w^\mathrm{error}}
\newcommand{\wscl}{w^\mathrm{scale}}
\newcommand{\wshp}{w^\mathrm{shape}}
\newcommand{\Werr}{W^\mathrm{error}}
\newcommand{\Wdist}{W^\mathrm{dist}}
\newcommand{\dscl}{\dl^\mathrm{scale}}
\newcommand{\dshp}{\dl^\mathrm{shape}}
\newcommand{\por}[1]{\ps_{#1}}
\newcommand{\landpor}{\ovl{\ps}}
\newcommand{\init}[1]{\overset{\raisebox{-2pt}{$\scriptstyle\circ$}}{#1}}
\newcommand{\va}{\vec{a}}
\newcommand{\vb}{\vec{b}}
\newcommand{\vc}{\vec{c}}
\newcommand{\vd}{\vec{d}}
\newcommand{\vg}{\vec{g}}
\newcommand{\vn}{\vec{n}}
\newcommand{\vp}{\vec{p}}
\newcommand{\vs}{\vec{s}}
\newcommand{\vv}{\vec{v}}
\newcommand{\iva}{\init{\va}}
\newcommand{\ivb}{\init{\vb}}
\newcommand{\ivc}{\init{\vc}}
\newcommand{\im}{\init{m}}
\newcommand{\imu}{\init{\mu}}
\newcommand{\iG}{\init{G}}
\newcommand{\iGinv}{\iG^{-1}}
\newcommand{\iVerts}{\init{\Verts}}
\newcommand{\iB}{{\init{\scr{B}}}}
\newcommand{\atp}{\va^\mathrm{tp}}
\newcommand{\btp}{\vb^\mathrm{tp}}
\newcommand{\ctp}{\vc^\mathrm{tp}}

\newcommand{\Sphere}{\mathrm{S}^2}
\newcommand{\area}{\operatorname{area}}
\newcommand{\caseselse}{\mathrm{else}}
\newcommand{\del}{\nabla}

\title{Minimum-Distortion Continuous Cartograms by Numerically Optimized Meshes}
\author{Robert C.\ Sargent}
\date{November 26, 2024}

\begin{document}

\maketitle

\interfootnotelinepenalty=10000

\begin{abstract}
    We present an algorithm for creating contiguous cartograms using meshes. We use numerical optimization to minimize cartographic error and distortion by transforming the mesh vertices. The vertices can either be optimized in the plane or optimized on the unit sphere and subsequently projected to the plane. We also present a hybrid ``best of both worlds'' method, where the vertices are optimized on the sphere while anticipating the distortion caused by the final projection to the plane. We show a significant improvement in the preservation of region shapes compared to existing automated methods. Outside the realm of cartograms, we apply this hybrid technique to optimized map projections, creating the Liquid Earth projection.
\end{abstract}

Thanks to Dr.\ Lawrence Washington for mentorship, encouragement, and feedback. Thanks to Justin Kunimune for encouragement, feedback, and programming help.

\section{Introduction}
Cartograms are data visualizations where each data point is represented as the area of a region. There are many different types of cartograms and algorithms for creating them, including manual methods. (For a comprehensive list of cartogram types and algorithms, see \cite{cart}.) This paper presents an algorithm for generating \emph{continuous} cartograms (also called \emph{contiguous} or \emph{deformation} cartograms), which continuously deform an existing map so that the region areas match the data points. We focus on world maps of countries with area representing population, but the techniques in this paper are applicable to cartograms of subnational divisions as well, and with areas representing any desired data.

\begin{figure}[H]
    \centering
    \includegraphics[width=0.8\linewidth]{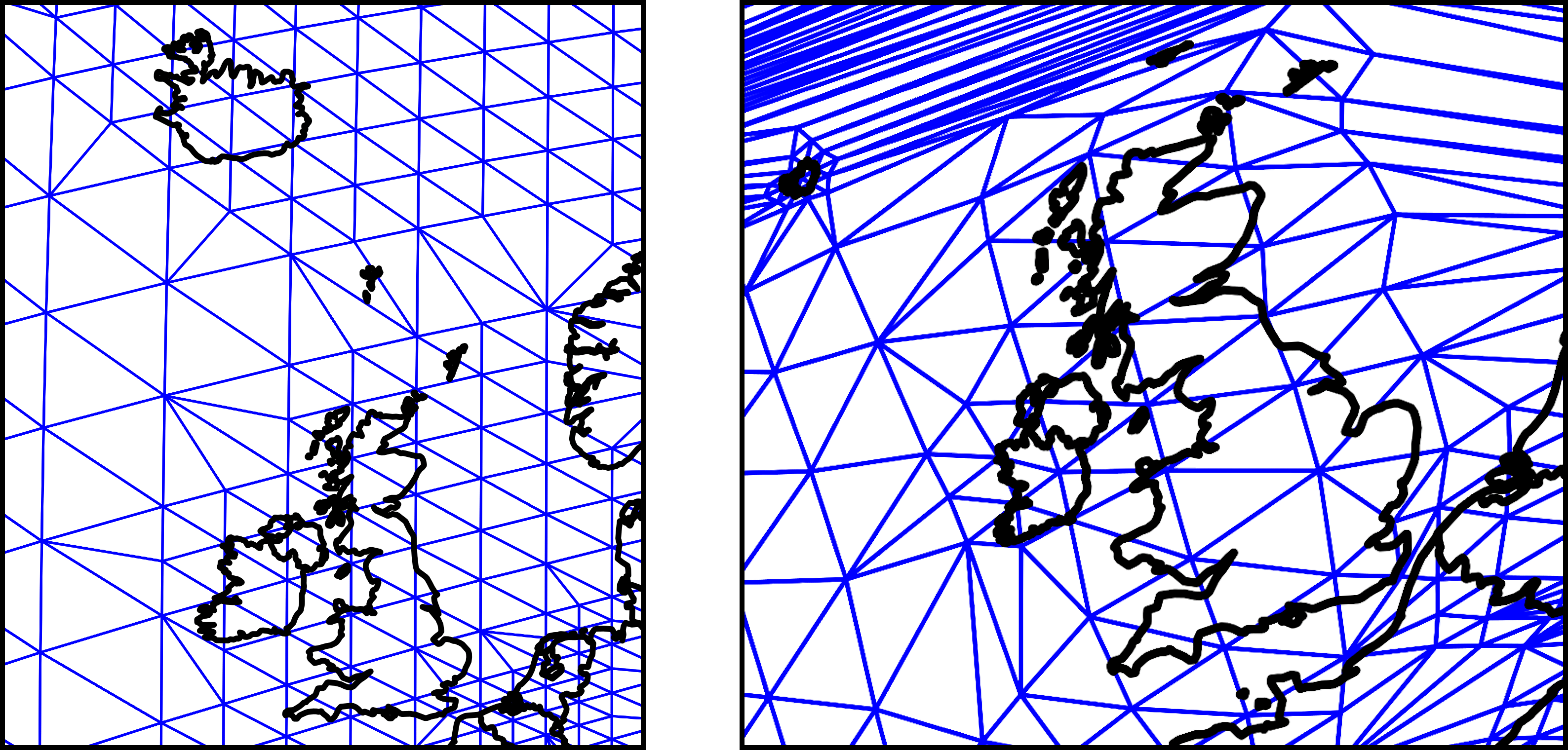}
    \caption{Part of the initial mesh and transformed mesh. Notice how the movement of the mesh vertices shrinks Ireland and Iceland.}
    \label{mesh-init-trans}
\end{figure}

Our algorithm works by projecting the globe onto a mesh of triangles. We project the globe onto what we call the \emph{initial mesh}, then transform the map by moving the mesh vertices to those of the \emph{transformed mesh} (\cref{mesh-init-trans}). We generate the transformed mesh so that the end result is a cartogram with minimal distortion.

To do this, we use numerical optimization, that is, finding a minimum of a multivariable function. We define a cost function whose inputs are the vertices of the transformed mesh. The cost function quantifies the total cartographic error (differences between the areas of regions and their desired values) and the total distortion. A local minimum of this function is an accurate cartogram with minimal distortion. The definition of this function depends on which version of the cartogram algorithm we use, as well as several hand-chosen parameters. For example, we choose to prioritize distortion of land much higher than distortion of water; we do this by giving the distortion of land triangles higher weight in the cost function.

Two key existing algorithms for automatic cartogram creation are the diffusion method of Gastner and Newman \cite{gn} and the rubber sheet method of Dougenik, Chrisman, and Niemeyer \cite{dcn}. The diffusion method works by imagining the map is filled by a fluid whose density at each point corresponds to population density, then letting the fluid diffuse to equal density. The rubber sheet method applies expansion and contraction functions to the plane with effects centered at each region, repeating the process a small number of times until the cartogram is accurate.

\begin{figure}[H]
    \centering
    \includegraphics[width=0.6\linewidth]{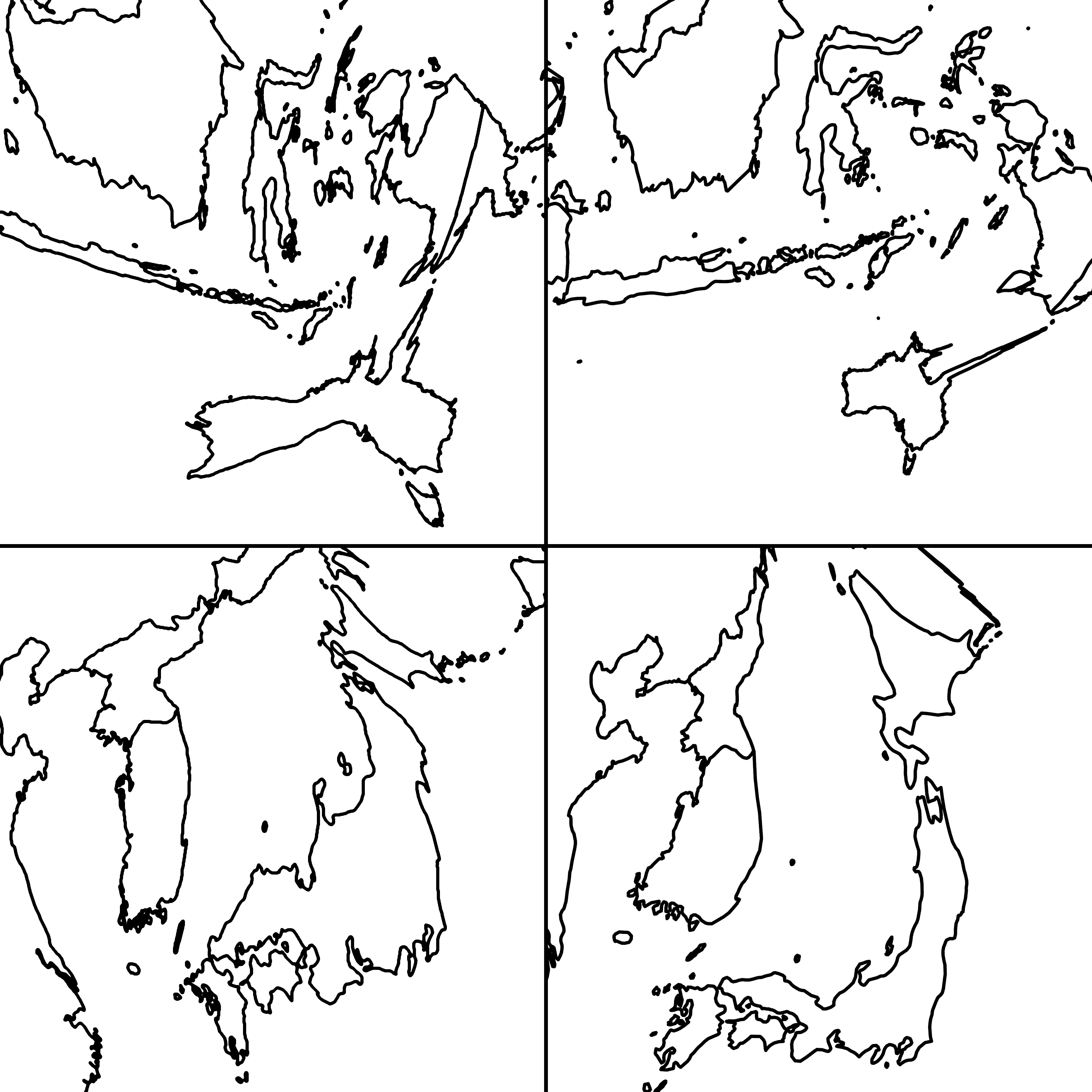}
    \caption{Comparison of shape preservation between the rubber-sheet method (left) and our mesh method (right). The mesh method preserves shapes as much as possible while preserving the connections (or near-connections) to adjacent landmasses. Rubber-sheet examples created with the program F4Carto by Sun \cite{odcn-c}.}
    \label{comparison}
\end{figure}

Though these methods can produce accurate cartograms, the resulting maps have more distortion than necessary. For example, an island country can be given the desired area by scaling the region up or down, without distorting its shape at all. However, existing algorithms cannot avoid introducing distortion in this situation. Because the method we present is specifically based on minimizing distortion, it is capable of avoiding distortion in this and other situations (\cref{comparison}).

Another issue with existing methods is that, with the exception of \cite{sphere}, they act only on the plane, not the sphere. This means that the globe must be projected to the plane before running the cartogram algorithm. This projection step bakes in distortion that the cartogram algorithm cannot correct. By contrast, our algorithm uses a spherical representation of the globe to calculate distortion, avoiding this problem. Li and Aryana \cite{sphere} present a diffusion algorithm that acts on the sphere, producing a cartogram on a globe. However, a final projection step is needed to produce a planar cartogram, which introduces distortion.

Though our algorithm significantly improves on these points, it has a clear disadvantage in runtime. Though the aforementioned algorithms take only seconds or minutes to run \cite{quant}, ours can take multiple hours. This is a hit to the practicality of our method. Optimizing the algorithm further is a potential area of future research.

\begin{figure}[H]
    \centering
    \includegraphics[width=0.8\linewidth]{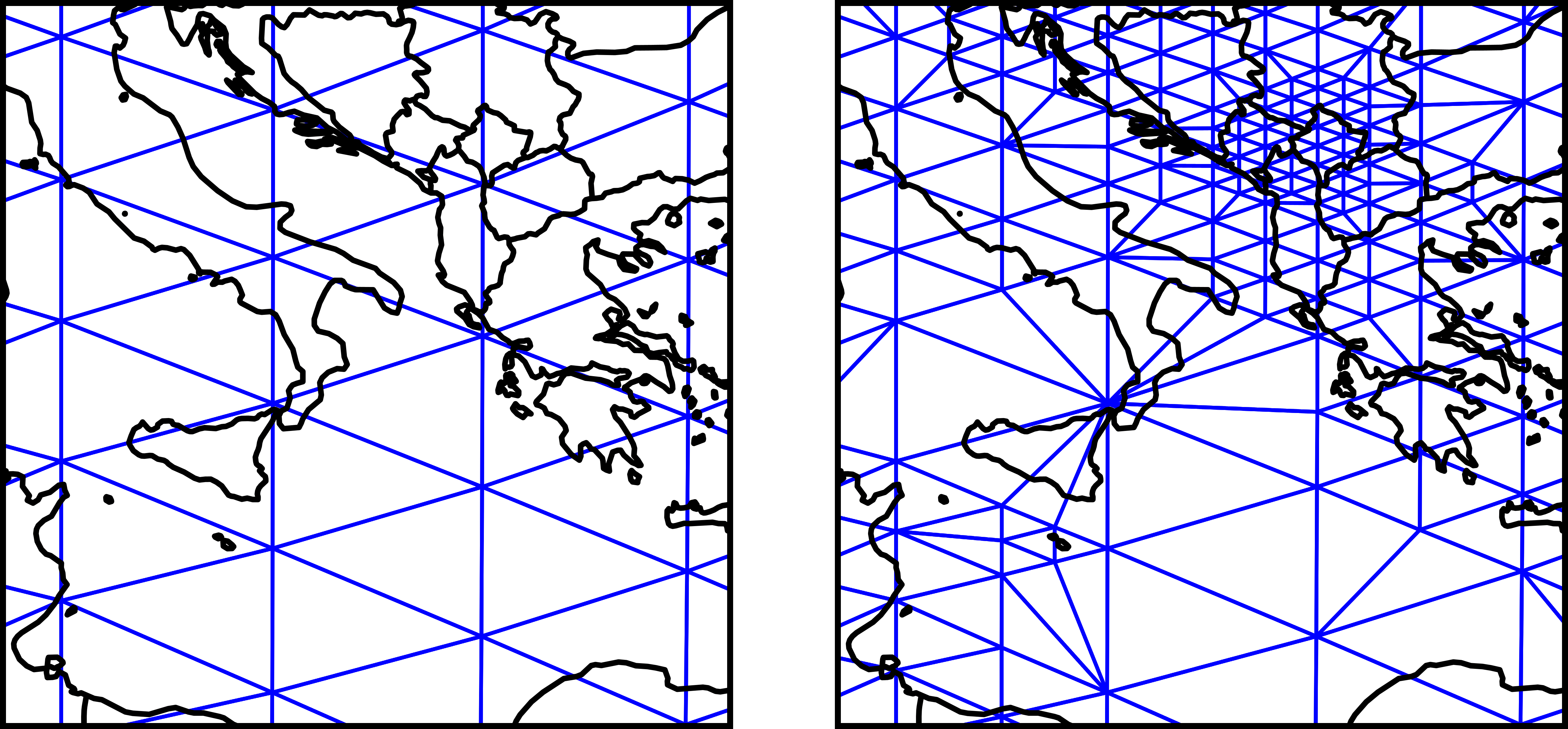}
    \caption{The initial mesh before and after subdivision.}
    \label{subdiv}
\end{figure}

The use of numerically optimized meshes for cartography has been explored by Kunimune (\cite{kun}, \cite{elastic}) and by Loncaric \cite{gatt} in the context of map projections. Kronenfeld \cite{man} develops a method for \emph{manually constructing} mesh-based cartograms. His method differs from ours in that, while we fix an initial mesh and optimize the transformed mesh, he fixes a transformed mesh and edits the initial mesh. This approach is helpful for dealing with small, population-dense regions (\cite{man} pp.\ 81--82). However, if we varied the initial mesh as he does, our cost function would not be differentiable, which would make numerical optimization infeasible. Instead, we deal with small, dense regions by adding more detail to the initial mesh before generating the cartogram (\cref{subdiv}).

\begin{figure}[H]
    \centering
    \includegraphics[width=1\linewidth]{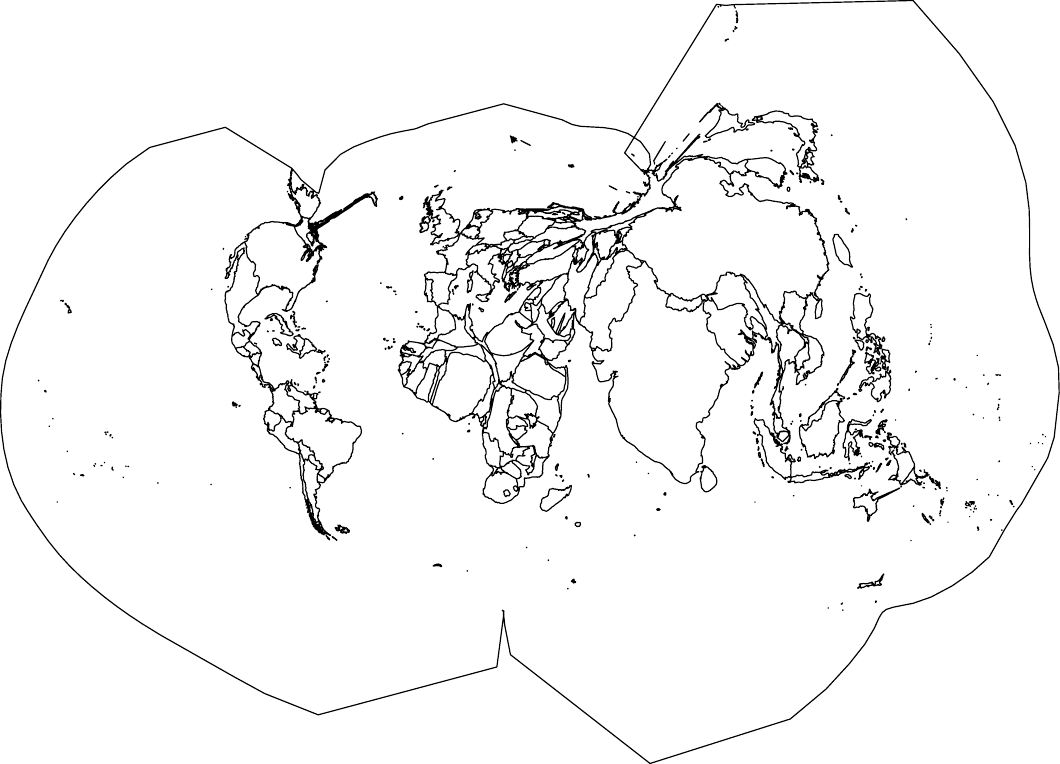}
    \caption{A plane cartogram.}
    \label{plane0}
\end{figure}

We present three different versions of our cartogram algorithm. The first is the \emph{plane cartogram} (\cref{plane0}), in which the mesh vertices are projected to the plane before the cost function is minimized. Because the mesh is cut before the optimization happens, the boundary of the final map is irregular, and there may be self-intersections at the boundary. This method is the most similar to \cite{kun}.

\begin{figure}[H]
    \centering
    \includegraphics[width=1\linewidth]{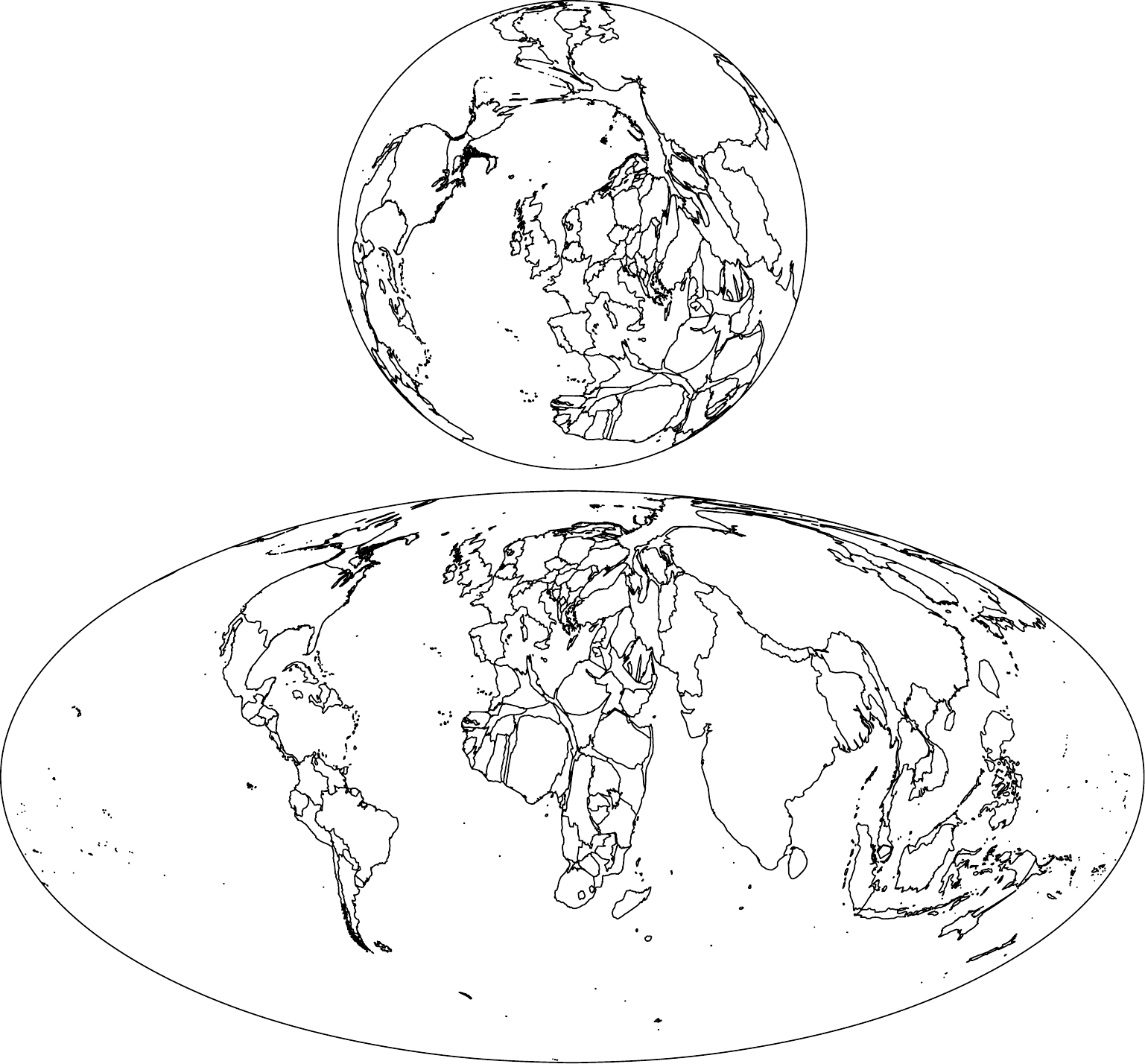}
    \caption{A sphere cartogram shown in orthographic view and projected using the Mollweide projection.}
    \label{sphere0}
\end{figure}

The second version is the \emph{sphere cartogram} (\cref{sphere0}). For this cartogram, the cost function takes in the positions of vertices on the unit sphere. Minimizing the cost function yields an accurate cartogram on the sphere. Since viewing a cartogram on a sphere is impractical, we then project the sphere to the plane using an equal-area map projection. The final projection step causes shape distortion. Since we optimize the cartogram on the sphere, topological problems are impossible. The boundary of the final map is regular; its shape is a consequence of the projection chosen.

\begin{figure}[H]
    \centering
    \includegraphics[width=1\linewidth]{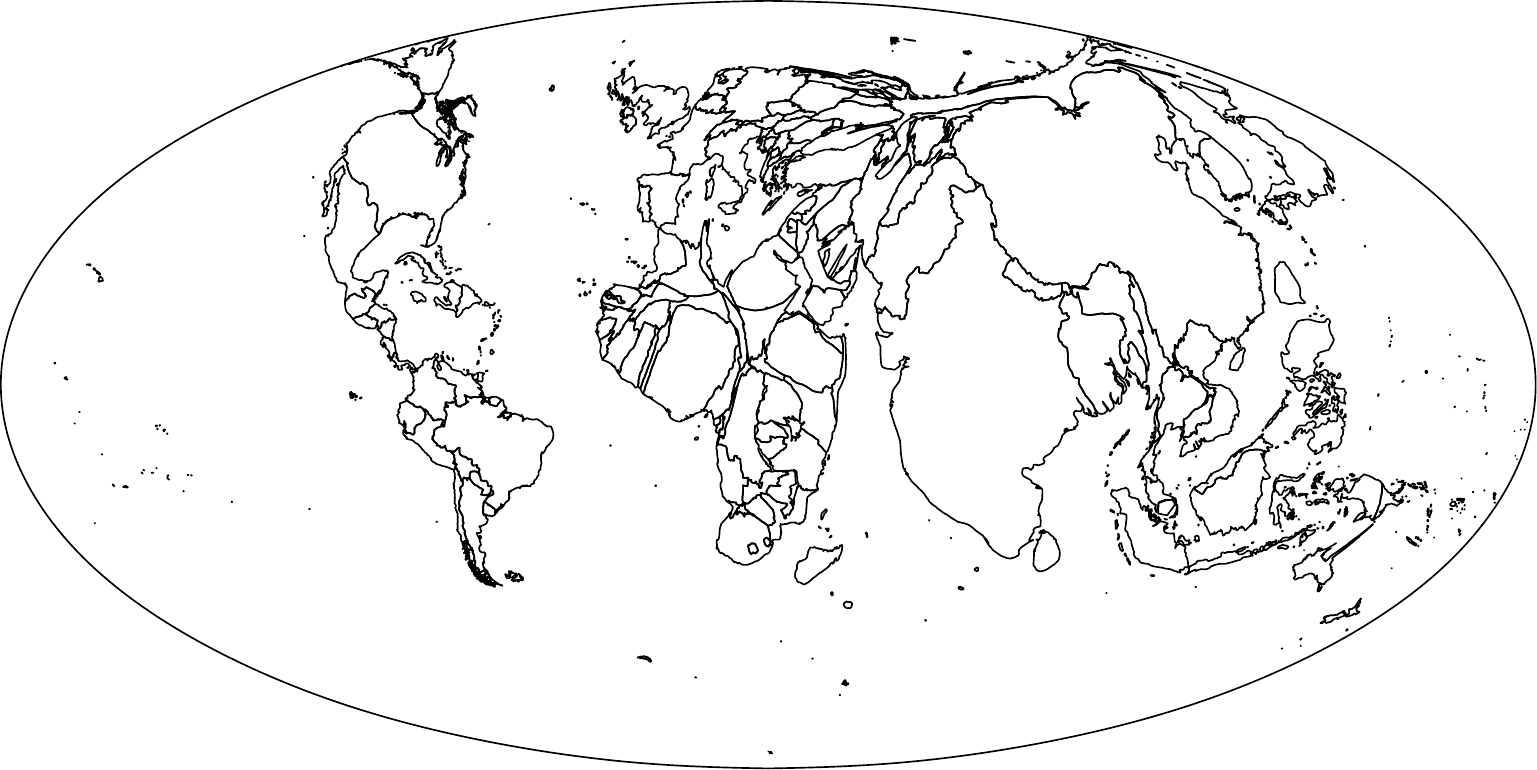}
    \caption{A hybrid cartogram whose target projection is Mollweide.}
    \label{hybrid0}
\end{figure}

Finally, the third version is the \emph{hybrid cartogram} (\cref{hybrid0}). Similarly to the sphere cartogram, it minimizes the cost function on the sphere, then projects the result to the plane. However, instead of calculating the distortion on the sphere, it anticipates how the mesh triangles will be transformed by the final projection and calculates distortion based on that. Minimizing this distortion function creates a cartogram that has minimal distortion \emph{after projection to the plane}. As a result, the hybrid cartogram has the low distortion of the plane cartogram, while keeping the topological soundness and clean boundary of the sphere cartogram. We call the final projection from the sphere to the plane the \emph{target projection}, because the optimization is done with this projection in mind.

\begin{figure}[H]
    \centering
    \includegraphics[width=1\linewidth]{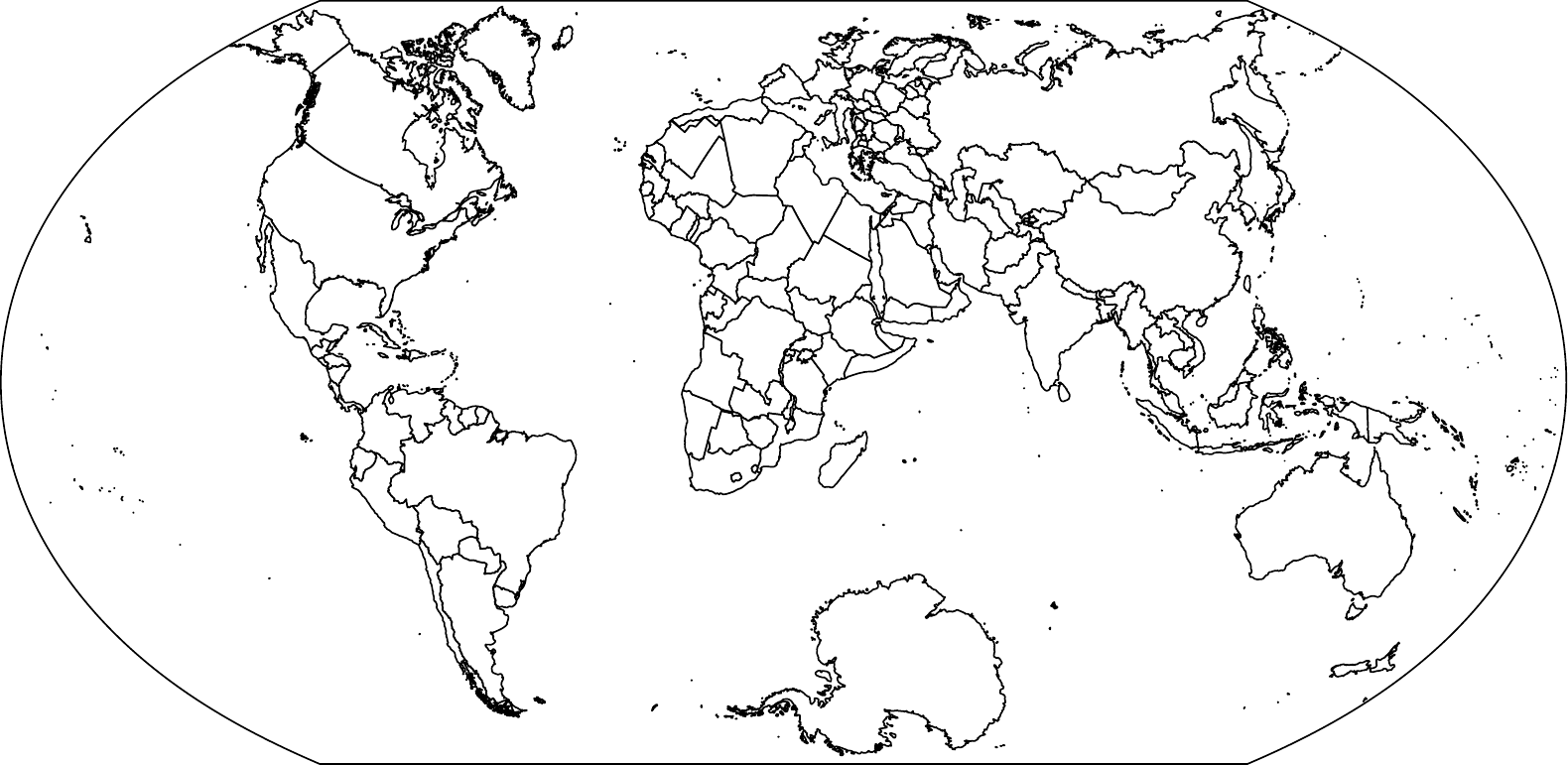}
    \caption{The Liquid Earth projection, an approximately equal-area map projection created by the hybrid technique. The target projection is Equal Earth.}
    \label{hybrid1}
\end{figure}

The technique used to create hybrid cartograms is also useful for creating new map projections. Kunimune (\cite{kun}, \cite{elastic}) prioritizes land distortion over water distortion to create some of the Danseiji and Elastic projections, as we do in our cartograms. However, he does his optimization on the plane. As with the hybrid cartogram, we can optimize a map projection on the sphere while anticipating the distortion caused by the final projection to the plane. This enables us to create a map projection that has exceptionally low distortion on land and whose boundary has a regular, familiar shape. \Cref{hybrid1} is a particularly good result of this process; I have released it independently as the Liquid Earth projection \cite{liquid}.

\Cref{overview} gives a high-level overview of the algorithm we present. \Cref{plane} gives a complete definition of the cost function used to create the plane cartogram, and shows how to compute its gradient. The cost functions for the sphere and hybrid cartograms are modifications of the one for the plane cartogram; we describe these modifications in \cref{sphere,hybrid}. \Cref{performance} quantifies the accuracy and runtime of the algorithm. We give final remarks in \cref{remarks}.

The source code for our algorithm is available at \url{https://github.com/rsargentmath/cartogram-gradient-descent}.

\section{Algorithm overview}\label{overview}
Throughout this paper, we use a right-handed coordinate system. Anticlockwise is the positive direction for both angles and polygons, including mesh triangles. The globe is always represented as the unit sphere, with the North Pole at $(0,0,1)$.

A cartogram is based on a set of \emph{regions} $\Regs$, in practice either countries or subnational divisions. Each region is a nonempty set of polygons on the unit sphere $\Sphere \sst \R^3$. (None of these polygons overlap, whether they're in the same region or different regions.) The boundaries of these polygons are called \emph{borders}. Each region $R \in \Regs$ has a number $r_R > 0$ associated with it, e.g., that country's population. These are the data the cartogram is trying to visualize. To be usable in the cartogram algorithm, we need to translate these values into areas we want the transformed regions to have. We define the desired area of each region by scaling the values $r_R$ so that the total of the desired areas equals the total of the initial areas:
\begin{equation*}
    p_R = r_R\, \frac{\sum_{S \in \Regs} \imu_S}{\sum_{S \in \Regs} r_S}\tc 
\end{equation*}
where $\imu_S$ is the initial area of $S$. In a slight abuse of terminology, we call the desired area $p_R$ the \emph{population} of $R$.

A \emph{mesh} consists of an array of vertices $\Verts$ and an array of triangles $\Tris$. The triangles in $\Tris$ do not contain vertices, rather, they are triples of indices into $\Verts$. This enables each vertex to be shared between multiple triangles. The choice to use triangles for our meshes differs from Kunimune (\cite{kun}, \cite{elastic}), who uses quads. We use triangles because the affine function that maps an initial triangle onto a transformed triangle scales all regions within the triangle evenly. This means only the portion that each region takes up of each triangle is needed to calculate the areas of transformed regions, as described below.

To create the specific mesh used in this paper, we start with a regular octahedron, then divide each face into $1024$ triangles of roughly equal size, resulting in a mesh of $8192$ triangles. After that, we subdivide the mesh further based on the borders and populations. We subdivide so that each region is covered by at least four triangles, and so triangles are expanded to at most $1/2048$ of the sphere's area. We also subdivide once at international borders and near the North Pole. The final mesh has $15396$ triangles. The choice of an octahedron makes it easy to interrupt the map along a meridian, where two opposing vertices of the octahedron correspond to the poles.

The algorithm starts with a mesh $\bigrou{\iVerts, \Tris}$ on the sphere, representing the undistorted initial state of the globe. We project the borders onto this mesh radially. (Though this projection slightly distorts area, no cartographic error can be introduced here, because both the area calculation and the final map are based on these projected borders.) The algorithm outputs a new array of vertices $\Verts$, forming a mesh $(\Verts, \Tris)$ where each triangle from the original mesh now references the new vertices. The final cartogram is created by projecting the borders from the original mesh to the new one. For each triangle, we use the unique affine transformation from the old triangle to the new one.

We want this process to result in an accurate cartogram that has minimal distortion. To achieve this, we use numerical optimization applied to a cost function $\Cost$ whose input is an array of vertices $\Verts$. This function incorporates the \emph{cartographic errors} (differences between the transformed areas of regions and their desired areas) as well as the total distortion of the map, which can be defined in various ways. The final vertices $\Verts$ are a local minimum of the function $\Cost$. By adjusting the definition of $\Cost$, we have fine control of the properties of the final cartogram. For example, we can weight scale distortion more or less heavily, and weight distortion of land more heavily than distortion of water.

To minimize $\Cost(\Verts)$, we start with an initial state $\Verts_0$, then step iteratively in the direction of lower cost until a local minimum is reached. For the sphere and hybrid cartograms, we simply set $\Verts_0 = \iVerts$. For the plane cartogram, this initial state needs to be in the plane. To do this, we modify the initial mesh $\bigrou{\iVerts, \Tris}$ so that the vertices on the antimeridian (excluding poles) are duplicated, then create $\Verts_0$ by projecting these vertices to the plane, using a projection that interrupts the antimeridian. Each pair of duplicated vertices in $\iVerts$ is separated by the interruption; these vertices form the outer edge of $\Verts_0$.

Suppose $n$ is the number of vertices in $\Verts$ and $d$ is the dimension of each vertex ($2$ or $3$). For the purpose of minimizing $\Cost(\Verts)$, we view $\Verts$ as a single vector in $\R^{nd}$ instead of a tuple of vectors in $\R^d$. With this view, the gradient vector $\del\Cost(\Verts)$ is also an element of $\R^{nd}$. The minimization algorithm we choose is L-BFGS \cite{l-bfgs} with backtracking line search \cite{line}, as was used by Kunimune \cite{kun}. This is a standard algorithm that works by computing and storing the gradient at each step, and using the gradients at the last few steps to choose the next step. The line search tries this step and computes its cost. If the cost does not decrease enough, or increases, it rejects the step and tries a smaller step. In particular, the line search always rejects a step if it leads to infinite cost. 

In order to avoid states with infinite cost, $\Cost$ must be continuous in the following sense. For any input $\Verts^*$ to $\Cost$, we must have $\lim_{\Verts\to\Verts^*} \Cost(\Verts) = \Cost(\Verts^*)$, including in the case where $\Cost(\Verts^*) = \infty$. If this fails and there are points where $\Cost(\Verts)$ discontinuously jumps to $\infty$, then the optimization algorithm can get stuck, because it has no way of steering around the disallowed states, despite knowing not to accept them.

Let $\Err(\Verts)$ be the total cartographic error and let $\Dist(\Verts)$ be the total distortion. The definitions of $\Err$ and $\Dist$ depend on both the type of the cartogram (plane, sphere, or hybrid) and various chosen weights, as described in this and following sections. We define maps with topological problems, such as having a triangle flipped over, to have infinite distortion; this ensures that the final output will not have these problems. The goal is to have $\Err(\Verts)$ as close to zero as possible (so that the cartogram is accurate) while minimizing $\Dist(\Verts)$ subject to this constraint. Achieving this is complicated by the fact that the initial state of the map does not satisfy $\Err(\Verts) = 0$. 

Our algorithm works by setting
\begin{equation*}
    \Cost(\Verts) = \Werr\,\Err(\Verts) + \Wdist\,\Dist(\Verts)\tc
\end{equation*}
where $\Werr, \Wdist \in \R$ are positive weights. We minimize $\Cost(\Verts)$ multiple times, each time prioritizing cartographic error higher by lowering $\Wdist$. The output of one step becomes the starting point of the next step's minimization. This way, the total cartographic error gets closer and closer to zero while distortion is still minimized. This process is reminiscent of the interior-point method from numerical optimization. (We cannot set $\Wdist$ to zero because $\Dist$ is responsible for maintaining the topology of the map.)

We define $\Err$ as follows. Let $\mu_R$ be the transformed area of a region $R$, which depends on the transformed vertices $\Verts$. (The method for computing $\mu_R$ depends on the type of the cartogram.) The cartographic error of $R$ is $\ep_R = \mu_R - p_R$. We define
\begin{equation*}
    \Err(\Verts) = \sum_{R \in \Regs} \werr_R \ep_R^2\tc
\end{equation*}
where $(\werr_R)_{R \in \Regs}$ is a family of positive weights. The function $\Err$ attains a minimum of $0$ precisely when $\mu_R = p_R$ for all regions $R$, as desired. These weights allow us to choose between minimizing absolute error, given by $\mu_R - p_R$, and relative error, given by $(\mu_R - p_R)/p_R$. If $\werr_R = 1/p_R^2$, then $\Err(\Verts)$ is the summed square relative error. For the cartograms in this paper, we split the difference and choose $\werr_R = 1/p_R$, since the extreme values given by $\werr_R = 1/p_R^2$ cause floating-point issues.

The total distortion $\Dist(\Verts)$ is computed by summing the scale distortion and shape distortion over all triangles $T \in \Tris$. As with cartographic error, we include arbitrarily chosen weights. This function may also include terms unrelated to these two types of distortion. For example, the plane cartogram described in \cref{plane} includes terms that prevent the map from having self-intersections at the poles. We give definitions for the scale and shape distortions at a triangle based on cartogram type (plane, sphere, or hybrid) in the following sections.


\section{Cost function for plane cartograms}\label{plane}
In this section, we fully define the cost function for the plane cartogram and show how to compute its gradient. The cost and gradient calculations for the sphere and hybrid cartograms are based on those for the plane, with the necessary modifications.

For each triangle $T$ and region $R$, we let $\por{R,T}$ be the portion $R$ takes up of $T$, that is,
\begin{equation*}
    \por{R,T} = \frac{\area(R \cap T)}{\area(T)}\tp
\end{equation*}
Since the maps from the initial mesh triangles to the transformed triangles are affine, the values $\por{R,T}$ do not depend on $\Verts$. These constants are all we need to compute the transformed area of each region; there is no complicated dependence on the geometry of the regions. We also compute the portion \emph{land} takes up of each triangle $T$, given by $\landpor_T = \sum_{R \in \Regs} \por{R,T}$.

For each triangle $T$, let $\iva_T, \ivb_T, \ivc_T \in \iVerts$ be the untransformed positions of the three vertices referenced by $T$, and let $\va_T, \vb_T, \vc_T \in \Verts$ be their transformed positions. For simplicity, we suppress $T$ and write $\iva, \ivb, \ivc, \va, \vb, \vc$. To define the distortion at $T$, we need to define the distortion of the affine transformation that takes $\iva, \ivb, \ivc$ to $\va, \vb, \vc$ respectively. Note that $\iva, \ivb, \ivc \in \R^3$ while $\va, \vb, \vc \in \R^2$. Let $\iB$ be an orthonormal basis for the plane parallel to the triangle $\iva, \ivb, \ivc$. Then we can analyze this transformation by looking at the $2\times2$ matrix $K$ taking $(\ivb - \iva)_\iB,\, (\ivc - \iva)_\iB$ to $\vb - \va,\, \vc - \va$ respectively. This matrix is given by
\begin{equation}\label{K}
    K = G\iGinv\tc
\end{equation}
where
\begin{equation}\label{GiG}
    \iG = \small\begin{bmatrix}
        | & | \\
        (\ivb - \iva)_\iB & (\ivc - \iva)_\iB \\
        | & |
    \end{bmatrix},\;\;\;\;
    G = \small\begin{bmatrix}
        | & | \\
        \vb - \va & \vc - \va \\
        | & |
    \end{bmatrix}\tp
\end{equation}
Since the distortion of a transformation should not be affected by rotation, the choice of $\iB$ does not matter as long as it is correctly oriented (in the sense that it gives $\iG$ a positive determinant). These matrices also allow us to compute the initial area $\im$ and transformed area $m$ of $T$:
\begin{equation}\label{imm}
    \im = \frac{1}{2}\det \iG,\;\;\;\; m = \frac{1}{2}\det G =  \im \,\det K\tp
\end{equation}

There are many ways to define the distortion of a $2\times2$ matrix. The definition chosen by Kunimune \cite{kun} has two problems for use in cartograms: (a) one term represents both scale distortion and shape distortion, when we need these to be separate; (b) the scale distortion term, which is the squared log of the scale factor, does not adequately punish the extreme scale distortions that can arise in cartograms. We give a definition that fixes these problems and makes the gradient simple to compute.

Let $A = \small\begin{bmatrix}
    a & b \\ c & d
\end{bmatrix}$ be any real $2\times2$ matrix. We define the shape distortion of $A$ to be
\begin{equation*}
    \begin{cases}
        \dfrac{a^2 + b^2 + c^2 + d^2}{ad - bc} - 2 & ad - bc > 0 \\
        \infty & \caseselse\tp
    \end{cases}
\end{equation*}
(In practice, such inequality checks are always implemented with a tolerance to compensate for floating-point error, e.g., checking $ad-bc > 10^{-12}$ instead of $ad-bc > 0$. For simplicity, we ignore this throughout this paper.) To justify this definition, notice that this fraction is the square of the Frobenius norm of $A$ divided by the determinant of $A$. Both of these values are invariant when multiplying $A$ by a rotation matrix on the left or right side. Thus, if $R_2 \small\begin{bmatrix}
    \sg_1 & 0 \\ 0 & \sg_2
\end{bmatrix} R_1$ is a singular value decomposition of $A$ ($R_1,R_2$ rotation matrices), then the shape distortion of $A$ equals the shape distortion of $\small\begin{bmatrix}
    \sg_1 & 0 \\ 0 & \sg_2
\end{bmatrix}$. If $\det A = \sg_1 \sg_2 > 0$, then this equals
\begin{equation*}
    \dfrac{\sg_1^2 + \sg_2^2}{\sg_1\sg_2} - 2 = \frac{\sg_1}{\sg_2} + \frac{\sg_2}{\sg_1} - 2\tp
\end{equation*}
This value is minimized at zero precisely when $\sg_1 = \sg_2$, i.e., when $A$ is a multiple of a rotation matrix. The constant term has no effect on the algorithm's outcome; its only purpose is cosmetic, ensuring that the shape distortion of a conformal transformation is measured as $0$ instead of $2$. We denote the shape distortion of our matrix $K$ by $\dshp$.

Note that, while we used the singular values to explain this definition, they are not necessary for computing this value. The distortion is given simply in terms of the matrix entries, which makes cost and gradient calculation easier. (\cite{gatt} defined the same distortion function, but gave it only in terms of the singular values.) An important characteristic is that, when fixing $\sg_1$, this function is convex in $\sg_2$, and vice versa. This property is important to prevent ``necking'' artifacts in the final map, where some triangles are near-perfect while others are extremely stretched \cite{kun}.

We define the scale distortion at this triangle in a similar way. We choose a value $s$ to be the \emph{intended scale} for this triangle. Then we let the scale distortion be
\begin{equation*}
    \dscl = \begin{cases}
        \dfrac{\det K}{s} + \dfrac{s}{\det K} - 2 & \det K > 0 \\
        \infty & \caseselse\tp
    \end{cases}
\end{equation*}
This is minimized at zero precisely when $\det K = s$. Note that the inequality check is the same for both $\dshp$ and $\dscl$.

If a triangle $T$ is taken up entirely by one region $R$, then it's clear that the intended scale of $T$ should equal the desired scaling of $R$, i.e., $s_T = p_R / \imu_R$, where $\imu_R$ is the initial area of $R$. For any region $R$, we compute $\imu_R$ by
\begin{equation}\label{imu}
    \imu_R = \sum_{T\in\Tris} \por{R,T}\,\im_T\tp
\end{equation}
In the general case where $T$ intersects at least one region, possibly including water as well, we choose an average of these values, weighted by how much of each region appears in $T$:
\begin{equation*}
    s_T = \sum_{R \in \Regs} \frac{\por{R,T}}{\landpor_T}\cdot \frac{p_R}{\imu_R}\tp
\end{equation*}
The remaining case is when $T$ contains entirely water, i.e., when $\landpor_T = 0$. In this case, it is not immediately clear how to choose $s_T$. For the cartograms in this paper, we blur the intended scale values across the mesh so that triangles with $\landpor_T = 0$ take on the values of nearby triangles, leaving triangles with $\landpor_T > 0$ unaffected. This improves the regularity of the mesh near coastlines.

Using \cref{imm}, we can compute the transformed area $\mu_R$ of a region $R$, similar to \cref{imu}:
\begin{equation*}
    \mu_R = \sum_{T\in\Tris} \por{R,T}\,m_T = \sum_{T\in\Tris} \por{R,T}\,\im_T\, \det K_T\tp
\end{equation*}
Thus, the total cartographic error is
\begin{equation*}
    \Err(\Verts) = \sum_{R \in \Regs} \werr_R \ep_R^2\tc
\end{equation*}
where
\begin{equation*}
    \ep_R = \mu_R - p_R = \Bigsqu{\sum_{T\in\Tris} \por{R,T}\,\im_T\, \det K_T} - p_R\tp
\end{equation*}

We define the total distortion $\Dist(\Verts)$ by
\begin{equation}\label{totaldist}
    \Dist(\Verts) = \sum_{T \in \Tris} \im_T \bigrou{\wshp_T \dshp_T + \wscl_T \dscl_T}\tc
\end{equation}
where $(\wshp_T)_{T\in\Tris}$ and $(\wscl_T)_{T\in\Tris}$ are families of positive weights. We multiply by $\im_T$ because distortion should be punished more if it affects a larger area.

The choice of weights for distortion has a large effect on the look of the final cartogram. For our cartograms, we incorporate both the presence or absence of land and the intended scale into these weights. To do this, we define auxiliary weight values that we multiply to obtain the final weights. We weight triangles that are entirely water much less than triangles that contain land, defining
\begin{equation*}
    \al^\mathrm{land}_T = \begin{cases}
        1 & \landpor_T > 0 \\
        0.1 & \caseselse\tp
    \end{cases}
\end{equation*}
Triangles that are mostly water but contain some of a region border should still receive full weight, because they are key to maintaining the visual shapes of regions. A small but significant weight on entirely-water triangles is still necessary to maintain the relative positions of regions and to avoid topological problems.

In \cref{totaldist}, we multiply each triangle $T$'s distortion by $\im_T$, its initial area. This means that the distortion of a region is judged using that region's size on the globe, not its size on the final map. In practice, this causes low--population density regions to be privileged over dense ones, e.g., leaving Canada mostly undistorted while the US warps to accommodate it. It would make more sense to judge distortion based on a triangle's final area. However, we can't just replace $\im_T$ with $m_T$ in \cref{totaldist}, because that would incentivize high-distortion triangles to shrink! To get around this, we include a weight factor based on each triangle's intended scale $s_T$ instead. So that very low-density areas are not completely ignored, we set
\[\al^\mathrm{density}_T = 0.2 + 0.8s_T\tp\]

Lastly, we choose how we weight shape distortion and scale distortion overall. For our cartograms, we choose
\[\al^\mathrm{shape} = 0.5,\;\; \al^\mathrm{scale} = 0.2\tp\]
For unclear reasons, higher values of $\al^\mathrm{scale}$ cause much longer runtimes, as the optimization takes many more steps to find a local minimum. Lower values improve runtime, but cause aesthetic problems. The given values are a compromise.

We define the final weights by
\[\wshp_T = \al^\mathrm{shape}\, \al^\mathrm{land}_T\, \al^\mathrm{density}_T,\;\;\;\wscl_T = \al^\mathrm{scale}\, \al^\mathrm{land}_T\, \al^\mathrm{density}_T\tp\]
With this, we have given a complete definition of the error and distortion functions for the plane cartogram. However, using this definition unmodified for the plane cartogram fails to prevent self-intersections at the poles. This is not an issue for the sphere and hybrid cartograms, so to keep the discussion generally applicable, we momentarily ignore this problem and proceed with the gradient calculation.

Recall that $\Cost(\Verts) = \Werr\,\Err(\Verts) + \Wdist\,\Dist(\Verts)$, so
\begin{equation*}
    \del\Cost(\Verts) = \Werr\,\del\Err(\Verts) + \Wdist\,\del\Dist(\Verts)\tc
\end{equation*}
as long as $\Cost(\Verts) < \infty$. Expanding this,
\begin{equation*}
    \del\Err(\Verts) = \del\Bigsqu{\sum_{R \in \Regs} \werr_R \ep_R^2} = 2\sum_{R \in \Regs} \werr_R \ep_R\, \del\ep_R\tc
\end{equation*}
\begin{equation*}
    \del\ep_R = \del\mu_R = \del\Bigsqu{\sum_{T\in\Tris} \por{R,T}\,\im_T\, \det K_T} = \sum_{T\in\Tris} \por{R,T}\,\im_T\, \del(\det K_T)\tp
\end{equation*}
Rearranging,
\begin{align*}
    \del\Err(\Verts) &= 2\sum_{R \in \Regs} \werr_R \ep_R \Bigsqu{\sum_{T\in\Tris} \por{R,T}\,\im_T\, \del(\det K_T)} \\
    & = 2\sum_{T \in \Tris}\Bigsqu{\sum_{R \in \Regs} \werr_R \ep_R\, \por{R,T}} \im_T\, \del(\det K_T)\tp
\end{align*}
Furthermore,
\begin{align*}
    \del\Dist(\Verts) &= \del\Bigsqu{\sum_{T \in \Tris} \im_T \bigrou{\wshp_T \dshp_T + \wscl_T \dscl_T}} \\
    &= \sum_{T \in \Tris} \im_T \bigrou{\wshp_T\, \del\dshp_T + \wscl_T\, \del\dscl_T}\tc
\end{align*}
so
\begin{align*}
    \del\Cost(\Verts) = &\;2\Werr \sum_{T \in \Tris}\Bigsqu{\sum_{R \in \Regs} \werr_R \ep_R\, \por{R,T}} \im_T\, \del(\det K_T) \\
    &+ \Wdist \sum_{T \in \Tris} \im_T \bigrou{\wshp_T\, \del\dshp_T + \wscl_T\, \del\dscl_T}
\end{align*}
\begin{equation}\label{costgrad}\begin{aligned}
    = \sum_{T\in\Tris}\im_T\Bigg[&\,2\Werr \Bigsqu{\sum_{R \in \Regs} \werr_R \ep_R\, \por{R,T}}\del(\det K_T) \\
    &+ \Wdist \bigrou{\wshp_T\, \del\dshp_T + \wscl_T\, \del\dscl_T}\,\Bigg]\tp
\end{aligned}\end{equation}
Since $\dshp_T$, $\dscl_T$, and $(\det K_T)$ are defined in terms of the entries of $K_T$, their derivatives (with respect to each component of $\Verts$) are simple to compute from the entries of $K_T$ and its derivatives. Finally, when differentiating with respect to each component of $\Verts$,
\begin{equation*}
    K_T\prm = (G_T\, \iGinv_T)\prm = G_T\prm\,\iGinv_T\tc
\end{equation*}
since $\iG_T$ is constant. The matrix $G_T\prm$ is immediate from \cref{GiG}; for example, the derivative of $G_T$ with respect to the $x$ component of $\va_T$ is $\small\begin{bmatrix}
    -1 & -1 \\ 0 & 0
\end{bmatrix}$.

Each triangle is only affected by the three vertices it references. This means that each element of the outermost sum in \cref{costgrad} has at most six nonzero components, namely the derivatives with respect to the components of the three vertices of that triangle. When computing $\del\Cost(\Verts)$ in code, we compute these six derivatives for each triangle $T$, then place them at the vertex indices referenced by $T$.

This cost function definition and gradient computation also work for the sphere cartogram. As described in \cref{sphere}, the only changes necessary are (a) the dimension of the vectors in $\Verts$; (b) the definition of the matrices $K_T$. We now describe a modification necessary to prevent the plane cartogram from self-intersecting; this modification does not apply to the sphere or hybrid cartograms.

\begin{figure}[H]
    \centering
    \includegraphics[width=0.6\linewidth]{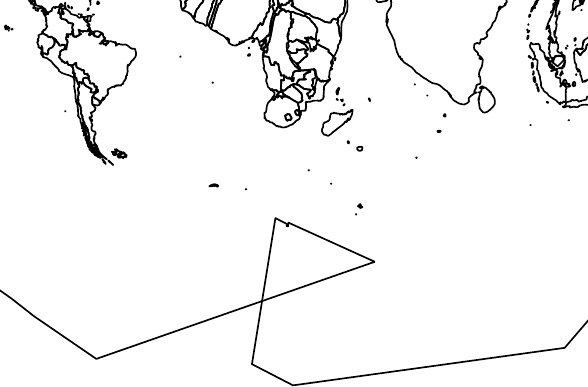}
    \caption{A large self-intersection caused by Antarctica contracting.}
    \label{self-inter}
\end{figure}

The self-intersection issue is worst at the South Pole, caused by Antarctica contracting. The land around the pole is shrank nearly to a point, while the surrounding water is shrank less. This situation makes the mesh behave like a negatively-curved surface, causing the angle at the pole to expand far past $360$ degrees (\cref{self-inter}). (Whether or not the cartogram actually represents country populations, Antarctica will likely have a data value near zero. This value cannot be exactly zero because that leads to division by zero. For our cartograms, we use an estimate of the temporary population of Antarctica, which has a small nonzero value.)

To prevent this problem, we add a term to the definition of $\Dist(\Verts)$ that compares the $x$ coordinates of vertices on the boundary to those of the poles. Points in the lower-right quadrant of the boundary are be required to be to the right of the South Pole, while those in the lower-left quadrant must be to its left. We do the same for the North Pole.

\begin{figure}[H]
    \centering
    \includegraphics[width=0.6\linewidth]{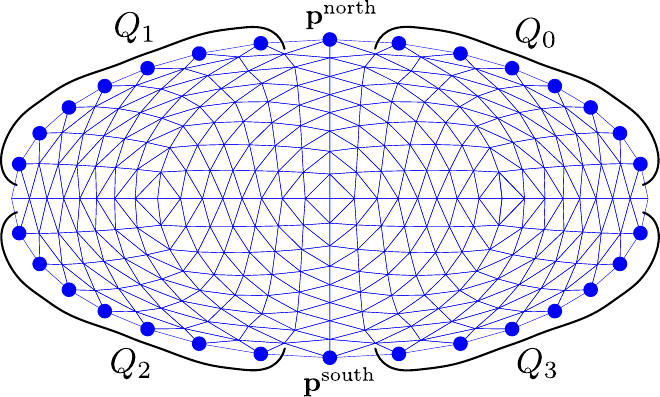}
    \caption{The sets of boundary vertices $Q_0, Q_1, Q_2, Q_3$ and the vertices $\vp^\mathrm{north}$ and $\vp^\mathrm{south}$. Mesh resolution reduced for clarity.}
    \label{quadrants}
\end{figure}

We define the sets $Q_0, Q_1, Q_2, Q_3 \sst \Verts$ of boundary vertices as in \cref{quadrants}, along with the poles $\vp^\mathrm{north}, \vp^\mathrm{south} \in \Verts$. We define a cost value for the boundary by
\begin{align*}
    \dl^\mathrm{boundary} = &\sum_{\vv \in Q_0} \frac{1}{\vv_x - \vp^\mathrm{north}_x} + \sum_{\vv \in Q_1} \frac{1}{\vp^\mathrm{north}_x - \vv_x} \\
    &+ \sum_{\vv \in Q_2} \frac{1}{\vp^\mathrm{south}_x - \vv_x} + \sum_{\vv \in Q_3} \frac{1}{\vv_x - \vp^\mathrm{south}_x}\tc
\end{align*}
with $\dl^\mathrm{boundary} := \infty$ if any of these denominators is not positive. We then modify the distortion function \cref{totaldist} to be
\begin{equation*}
    \Dist(\Verts) = \Bigsqu{\sum_{T \in \Tris} \im_T \bigrou{\wshp_T \dshp_T + \wscl_T \dscl_T}} + w^\mathrm{boundary}\dl^\mathrm{boundary}\tp
\end{equation*}
We set $w^\mathrm{boundary}$ to be quite small, specifically $10^{-6}$, so that it only has an effect when a boundary point is very close to crossing over. The corresponding modification to $\del\Dist(\Verts)$ is straightforward.

This addition does not prevent all self-intersections in the final map, but it avoids a common and severe case. The sphere and hybrid cartograms do guarantee a lack of topological problems by construction.

\section{Sphere cartograms}\label{sphere}
While the transformed vertices in the plane cartogram are in $\R^2$, the transformed vertices in the sphere cartogram are in $\Sphere \sst \R^3$. The borders in each triangle are mapped using the affine function taking the initial vertices $\iva,\ivb,\ivc$ to the transformed vertices $\va,\vb,\vc$, then projected radially onto $\Sphere$.

Care must be taken when adapting the cost function for the sphere. With this radial projection to the sphere, the maps from the initial triangles to the transformed triangles of the cartogram are no longer affine. In order to measure distortion using the methods in \cref{plane}, we need to choose affine approximations of these maps. The obvious choice is to ignore the projection to the sphere and just look at the map from $\iva,\ivb,\ivc$ to $\va,\vb,\vc$. However, this results in the cost function being discontinuous. Points that are collinear on the sphere are in general not collinear in $\R^3$, so if a triangle is in the process of flipping over, it maintains significant positive area right up until it flips. This causes the shape and scale distortion values to discontinuously jump to infinity. The option we choose instead is to project the transformed triangle onto the tangent plane at its midpoint, then calculate distortion based on this projected triangle.

Define the function $\nzd$ (``normalized'') by $\nzd(\vv) = \dfrac{\vv}{\Ver{\vv}}$. Let $\va,\vb,\vc$ be the transformed vertices of some fixed triangle. We define the midpoint of this triangle to be 
\begin{equation}\label{midpoint}
    \vn = \nzd\lrrou{\frac{\va+\vb+\vc}{3}} = \nzd(\va+\vb+\vc)\tp
\end{equation}
We project $\va,\vb,\vc$ perpendicularly onto the tangent plane to the sphere at $\vn$, yielding the points $\atp,\btp,\ctp$. The point $\atp$ is given by
\begin{equation*}
    \atp = \va + (1 - \va \cdot \vn)\,\vn,
\end{equation*}
with $\btp,\ctp$ given similarly. Finally, we modify the definition of the matrix $K$ [\cref{K,GiG}] by setting
\begin{equation}\label{Gsphere}
    G = \small\begin{bmatrix}
        | & | \\
        (\btp - \atp)_\scr{B} & (\ctp - \atp)_\scr{B} \\
        | & |
    \end{bmatrix}\tc
\end{equation}
where $\scr{B} = (\scr{B}_1, \scr{B}_2)$ is a correctly oriented orthonormal basis for the plane parallel to this tangent plane. (Since $\iG$ is constant, there is no need to modify its definition in a similar way.)

The area of the planar triangle $(\atp,\btp,\ctp)$ is not quite equal to the area of the spherical triangle $(\va,\vb,\vc)$, and the radial projection to the sphere is not area-equivalent, so a small amount of cartographic error is introduced. We estimate this error in \cref{performance}.

The basis $\scr{B}$ is dependent on the tangent plane at $\vn$, which depends on $\va,\vb,\vc$. Since $\scr{B}$ is not constant as the mesh vertices vary, we must give a fixed definition for it and pay attention to how it moves. We choose $\scr{B}$ to align with the longitude--latitude graticule, meaning that $\scr{B}_1$ points east and $\scr{B}_2$ points north. This choice is convenient when defining hybrid cartograms (\cref{hybrid}). (If $\vn$ is at a pole, we choose $\scr{B}$ arbitrarily.) With this choice made, it is possible to describe each component of $G$ directly in terms of $\va,\vb,\vc$, and use those formulas to calculate the gradient of the cost function. However, it is simpler to calculate the gradient with respect to $\atp,\btp,\ctp$, then use the chain rule to find the gradient in terms of $\va,\vb,\vc$, as described below.

Notice that, if $\atp$, $\btp$, or $\ctp$ were to vary in the direction perpendicular to the tangent plane, this would have no first-order effect on the distortion of the triangle $(\atp,\btp,\ctp)$. This means that, when computing the gradient, we only need to consider the movement of these points within the tangent plane. As for the basis, notice that as the midpoint $\vn$ varies in the east--west direction, $\scr{B}$ rotates so that the second basis vector points points towards the North Pole. This is a first-order effect, so it affects the derivatives of $G$. However, our area and distortion calculations are unaffected by rotation. Thus, when computing the gradient of the cost, we can view both the tangent plane and the basis as fixed. If the rotation of $G$ does matter, as in the case of the hybrid cartogram, we correct for this after the fact, as described in \cref{hybrid}.

Since we view $\atp,\btp,\ctp$ as varying within a fixed plane, the computations for the derivatives of $\det K$, $\dshp$, and $\dscl$---with respect to these vertices moving along $\scr{B}_1$ and $\scr{B}_2$---are the same as the corresponding computations for the plane cartogram. The resulting gradient vectors are expressed in the basis $\scr{B}$; we then use $\scr{B}$ to express them in global coordinates. Finally, we use the chain rule to find the derivatives of $\det K$, $\dshp$, and $\dscl$ with respect to the components of $\va$, $\vb$, and $\vc$. This involves finding the derivatives of $\atp,\btp,\ctp$ with respect to each of these components. Let $\vd = \va + \vb + \vc$. To find each of these derivatives, we compute
\begin{equation*}
    \vd\prm = \va\prm+\vb\prm+\vc\prm\tc
\end{equation*}
\begin{align*}
    \vn\prm &= \nzd(\vd)\prm = \lrrou{\frac{\vd}{\Ver{\vd}}}\prm = \bigrou{(\vd\cdot\vd)^{-1/2}\,\vd}\prm \\
    &= -\frac{1}{2}(\vd\cdot\vd)^{-3/2}\,2(\vd\cdot\vd\prm)\,\vd + (\vd\cdot\vd)^{-1/2}\,\vd\prm \\
    &= -(\vd\cdot\vd)^{-3/2}\,(\vd\cdot\vd\prm)\,\vd + (\vd\cdot\vd)^{-1/2}\,\vd\prm\tp
\end{align*}
Then
\begin{align*}
    (\atp)\prm &= \bigrou{\va + (1 - \va \cdot \vn)\,\vn}\prm \\
    &= \va\prm - (\va\prm\cdot\vn + \va\cdot\vn\prm)\,\vn + (1-\va\cdot\vn)\,\vn\prm\tc
\end{align*}
and similarly for $(\btp)\prm$ and $(\ctp)\prm$. Notice that each of $\atp,\btp,\ctp$ depends on all of $\va,\vb,\vc$.

It remains to discuss how to actually carry out this optimization when the mesh vertices' domain is the sphere. In our implementation, we represent the vertices with 3D Cartesian coordinates, as we do mathematically. After each step (in the search direction provided by L-BFGS), we normalize each vertex back onto the sphere. Explicitly, if $\Verts = (\vv_0,\dots,\vv_{n-1})$ is the current vertex array ($\vv_i \in \Sphere \sst \R^3$), and $\scr{S} = (\vs_0,\dots,\vs_{n-1})$ is the step produced by L-BFGS and the line search, then the next positions of the mesh vertices, $\Verts^\mathrm{next} = (\vv_0^\mathrm{next},\dots,\vv_{n-1}^\mathrm{next})$, is given by $\vv_i^\mathrm{next} = \nzd(\vv_i + \vs_i)$ for all $i$. This normalization step introduces a small challenge in the implementation.

The line search is a standard part of the optimization algorithm. The program tries the step suggested by L-BFGS and compares the resulting reduction in cost to the reduction that is expected based on the gradient. If the cost increases or the reduction is otherwise too small, we try again with a smaller step. This is repeated until an acceptable step is found. Viewing $\Verts$, $ \Verts^\mathrm{next}$, $\del\Cost(\Verts)$, and $\scr{S}$ as vectors in $\R^{3n}$, we define $\scr{S}$ to be acceptable if it satisfies Armijo's condition \cite{line}:
\begin{equation}\label{armijo}
    \Cost(\Verts^\mathrm{next}) - \Cost(\Verts) \leq c\bigrou{\scr{S} \cdot \del\Cost(\Verts)}\tc
\end{equation}
where $c \in (0, 1)$ is a constant (we choose $0.1$).

\begin{figure}[h]
    \centering
    \includegraphics[width=0.5\linewidth]{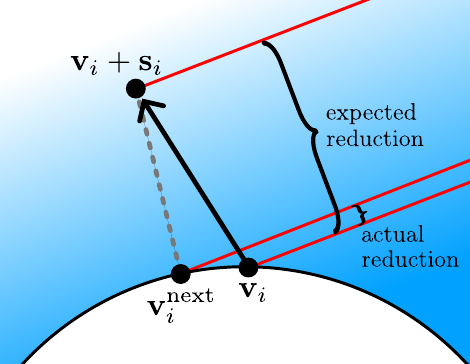}
    \caption{How normalizing vertices changes the cost reduction. The shading represents cost as $\vv_i$ varies, with blue representing higher cost.}
    \label{normalization}
\end{figure}

If there is no normalization, then we are guaranteed to find a step satisfying \cref{armijo}, since $\scr{S} \cdot \del\Cost(\Verts)$ is a first-order approximation for $\Cost(\Verts+\scr{S}) - \Cost(\Verts)$ as $\scr{S}$ varies, by the definition of the gradient. However, the normalization step causes a problem. The gradient $\del\Cost(\Verts)$ can direct points to move off the sphere, in which case the step $\scr{S}$ will likely move points off the sphere. When these points are normalized, the reduction in cost is smaller than is expected based on $\scr{S}$ (\cref{normalization}). If $\scr{S}$ and $\del\Cost(\Verts)$ have significant off-sphere components, then \cref{armijo} may not be satisfied no matter how small $\scr{S}$ is chosen to be. If this happens, the line search cannot find an acceptable step, and the algorithm fails.

To fix this problem, we modify the gradient before it's passed into the optimization algorithm. Write $\del\Cost(\Verts) = (\vg_0,\dots,\vg_{n-1})$. We remove the off-sphere component from each $\vg_i$, defining the modified gradient $\scr{G}(\Verts) = (\widetilde{\vg}_0,\dots,\widetilde{\vg}_{n-1})$ by
\begin{equation*}
    \widetilde{\vg}_i = \vg_i - (\vg_i \cdot \vv_i)\,\vv_i\tp
\end{equation*}
We then pass $\scr{G}(\Verts)$ into the optimization algorithm in place of $\del\Cost(\Verts)$. In particular, we use $\scr{G}(\Verts)$ in place of $\del\Cost(\Verts)$ in \cref{armijo}. Then the off-sphere component of $\scr{S}$ no longer contributes to the dot product in \cref{armijo}, so we can always satisfy the condition with a sufficiently small step. Modifying the gradient in this way means there is no need to modify the optimization algorithm, with the exception of including the normalization step.

If the reader wants to build their own implementation, and modifying the optimization algorithm to include this normalization step is not an option, a possible approach is to represent the mesh vertices with two coordinates each by parametrizing the sphere. Any parametrization of the sphere has at least one singularity, which could cause problems with this approach. One option is to choose a different parametrization for each vertex, so that the singularity for each vertex is opposite its original position, making it very unlikely for problems to arise.

\section{Hybrid cartograms}\label{hybrid}
To create the hybrid cartogram, we calculate distortion based on how a triangle is transformed after it is projected to the plane (using a fixed equal-area projection). To describe this, we need the Jacobian matrix of this map projection at each point on the sphere. Let $\om : \Sphere \to \R^2$ be the target projection. To represent the total derivative of $\om$ as a matrix, we parametrize $\Sphere$ using longitude and latitude, denoted $\lm$ and $\ph$ respectively. The point on $\Sphere$ referenced by $\lm,\ph$ is
\begin{equation*}
    \bigrou{{\cos\ph \cos\lm},\,\cos\ph\sin\lm,\, \sin\ph}\tp
\end{equation*}
Using this parametrization, we view $\om$ as a map $[-\pi, \pi]\times[-\pi/2,\pi/2] \to \R^2$. At each point $\vp \in \Sphere$, we choose the basis for the tangent space at $\vp$ to align with the graticule, as we did in the previous section. Then, writing the output of $\om$ as $(x,y)$, if $\vp$ has the coordinates $\lm,\ph$, then the Jacobian matrix of $\om$ at $\vp$ is given by
\begin{equation*}
    \mathrm{J}_\om(\vp) = 
    \begingroup
    \renewcommand*{\arraystretch}{2}
    \begin{bmatrix}
        \dfrac{1}{\cos\ph}\, \dpdpd{x}{\lm} & \dpdpd{x}{\ph} \\
        \dfrac{1}{\cos\ph}\, \dpdpd{y}{\lm} & \dpdpd{y}{\ph}
    \end{bmatrix}
    \endgroup\tp
\end{equation*}

For use in the cost function, we measure the Jacobian at the center point of each triangle. Unfortunately, $\mathrm{J}_\om(\vp)$ jumps discontinuously as $\vp$ crosses an interruption of the projection. In particular, for the pseudocylindrical projections we use, $\mathrm{J}_\om$ is discontinuous along the antimeridian. Since the cost function must be continuously differentiable, we blur $\mathrm{J}_\om$ near the interruption. Specifically, we choose a continuously differentiable function $H : \Sphere \to \R^{2\times2}$ such that $H(\vp) = \mathrm{J}_\om(\vp)$ unless $\vp$ is close to the interruption, and use $H(\vp)$ in place of $\mathrm{J}_\om(\vp)$ when calculating distortion.\footnote{Because we choose bases for the tangent spaces to align with the longitude/latitude graticule, and this graticule has singularities at the poles, it is not sufficient to simply have $H$ continuously differentiable as a function $\Sphere \to \R^{2\times2}$. The correct condition is as follows. Instead of viewing $H(\vp)$ as a member of $\R^{2\times2}$ for each $\vp$, we can say that $H(\vp)$ is a linear map $\mathrm{T}_\vp\Sphere \to \mathrm{T}_\vp\Sphere$, where $\mathrm{T}_\vp\Sphere$ is the tangent space at $\vp$. Suppose that $\scr{B}(\vp)$ is a basis for $\mathrm{T}_\vp\Sphere$ for each $\vp$. Then, using $\scr{B}(\vp)$, we can view each $H(\vp)$ as a linear map $H_\scr{B}(\vp) : \R^2 \to \R^2$. This creates a function $H_\scr{B} : \Sphere \to \R^{2\times2}$. The condition we need is that $H_\scr{B}$ must be continuously differentiable for \emph{any} smooth choice of bases $\scr{B}(\vp)$ (defined over any subset of $\Sphere$). In practice, we achieve this by choosing $H(\vp)$ to be the identity when $\vp$ is a pole.} We omit the specific definition of $H$. 

We only take $H$ into account when calculating shape distortion, leaving the scale distortion and area calculations unaffected. This means there is no need to ensure that $H(\vp)$ always has determinant $1$. (Since $\om$ is equal-area, $\mathrm{J}_\om(\vp)$ always has determinant $1$, but the blur can change this.) For each triangle, we multiply the output of $H$ by the matrix $K$ [\cref{K,GiG,Gsphere}], defining
\begin{equation*}
    \tl{K} = H(\vn)\,K\tc
\end{equation*}
where $\vn$ is the midpoint of the triangle, given by \cref{midpoint}. The matrix $\tl{K}$ approximately represents the map taking the initial triangle $(\iva,\ivb,\ivc)$ to the projected triangle $\bigrou{\om(\va),\,\om(\vb),\,\om(\vc)}$. We use $\tl{K}$ to calculate $\dshp$, with the scale distortion and area just using $K$ as before.

\begin{figure}[H]
    \centering
    \includegraphics[width=0.55\linewidth]{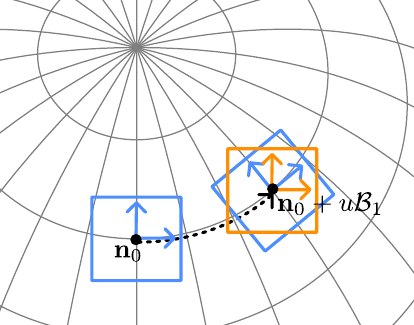}
    \caption{The effect on the local basis caused by moving east in the Northern Hemisphere. Orange represents the result of moving the basis at $\vn_0$ without rotating it, as we do in \cref{sphere}. From the perspective of the graticule-aligned basis (blue), moving east causes a clockwise rotation.}
    \label{basis_rotation}
\end{figure}

Recall that when $\vn$ moves in the east--west direction, the basis for the tangent space at $\vn$ rotates, and that we ignore this effect when doing the calculations in \cref{sphere}. The rotation of the basis has the effect of multiplying $K$ by a rotation matrix on its left (see \cref{basis_rotation}). Since $H(\vn)$ is subsequently multiplied on the left, this has a non-rotational effect on $\tl{K}$, so it must be taken into account when calculating the gradient.

For the purpose of finding derivatives, we call the current midpoint of the triangle $\vn_0$ and view the midpoint $\vn$ as varying in the tangent plane at $\vn_0$. We write $\vn = \vn_0 + u\scr{B}_1 + v\scr{B}_2$, $u,v\in\R$, where $\scr{B} = (\scr{B}_1,\scr{B}_2)$ is the orthonormal basis for the tangent space at $\vn_0$. Since the notational distinction between $\vn_0$ and $\vn$ is just for the purpose of taking derivatives, all of these derivatives are evaluated at $\vn = \vn_0$. We represent the induced rotation relative to the basis vectors by the rotation matrix $R(u,v)$. For example, in \cref{basis_rotation}, $R(u,v)$ is a clockwise rotation. The formula for $\tl{K}$ (for the purpose of finding derivatives) becomes
\begin{equation*}
    \tl{K} = H(\vn)\,R(u,v)\,K\tc
\end{equation*}
so each derivative of $\tl{K}$ (evaluated at $\vn = \vn_0$) is
\begin{align*}
    \tl{K}\prm &= [H(\vn)\,R(u,v)\,K]\prm\\
    &= [H(\vn)\,R(u,v)]\prm\,K + [H(\vn)\,R(u,v)]\,K\prm \\
    &= [H(\vn)\,R(u,v)]\prm\,K + [H(\vn_0)\,R(0,0)]\,K\prm \\
    &= [H(\vn)\,R(u,v)]\prm\,K + H(\vn_0)\,K\prm\tp
\end{align*}
The calculations for $K$ and $K\prm$ are covered in the previous section, so it only remains to find $[H(\vn)\,R(u,v)]\prm$.

It suffices to find the derivatives of $H(\vn)\,R(u,v)$ with respect to $u$ and $v$; the derivatives with respect to the components of the triangle's vertices are then found using the chain rule. We can write $H(\vp) = H(\lm,\ph)$ using the usual parametrization of the sphere. Let $\lm_0,\ph_0$ be the coordinates corresponding to $\vn_0$. The derivative with respect to $v$ is simple:
\begin{align*}
    \pdpd{}{v}[H(\vn)\,R(u,v)]\Big|_{\vn=\vn_0} &= \pdpd{H}{v}(\vn_0)\,R(0,0) + H(\vn_0)\,\pdpd{R}{v}(0,0) \\
    &= \pdpd{H}{v}(\vn_0)\,I + H(\vn_0)\,0 \\
    &= \pdpd{H}{v}(\vn_0) = \pdpd{H}{\ph}(\lm_0,\ph_0)\tc
\end{align*}
since varying $v$ (i.e., moving north--south) induces no rotation of the basis. Deriving with respect to $u$, we have
\begin{align}
    \pdpd{}{u}[H(\vn)\,R(u,v)]\Big|_{\vn=\vn_0} &= \pdpd{H}{u}(\vn_0)\,R(0,0) + H(\vn_0)\,\pdpd{R}{u}(0,0) \nonumber \\
    &= \pdpd{H}{u}(\vn_0)\,I + H(\vn_0)\,\pdpd{R}{u}(0,0) \nonumber \\
    &= \frac{1}{\cos \ph_0}\,\pdpd{H}{\lm}(\lm_0,\ph_0) + H(\lm_0,\ph_0)\,\pdpd{R}{u}(0,0)\tc \label{dduHR}
\end{align}
since a change in $u$ is $\cos \ph_0$ times the corresponding change in the longitude $\lm$.

It remains to find $\dpdpd{R}{u}(0,0)$. To find this, we view $\vn$ as varying along the latitude line at $\ph_0$. Having moved a distance of $u$ along this line, the basis $\scr{B}$ has rotated by some angle $\th(u)$. Then we have
\begin{equation*}
    R(u,0) = \begin{bmatrix}
        \cos(-\th(u)) & -\sin(-\th(u)) \\
        \sin(-\th(u)) & \cos(-\th(u))
    \end{bmatrix} = \begin{bmatrix}
        \cos\th(u) & \sin\th(u) \\
        -\sin\th(u) & \cos\th(u)
    \end{bmatrix}\tp
\end{equation*}
The negation appears because $R(u,0)$ represents the rotation \emph{from the perspective of} the basis. Then
\begin{equation}\label{dRdu}
    \dpdpd{R}{u}(0,0) = \pdpd{}{u}\begin{bmatrix}
        \cos\th(u) & \sin\th(u) \\
        -\sin\th(u) & \cos\th(u)
    \end{bmatrix}\Bigg|_{u=0} = \th\prm(0)\begin{bmatrix}
        0 & 1 \\ -1 & 0
    \end{bmatrix}\tp
\end{equation}

The value $\th\prm(0)$ is the radians of rotation of the basis per distance travelled along the latitude line. Since the basis is aligned with the line, this is the radians of rotation of the line per distance travelled. By definition, this is the geodesic curvature of this line. To find this value, we find the total geodesic curvature of the latitude line using the Gauss--Bonnet theorem. The area of the region of $\Sphere$ enclosed (to th\textbf{e} north) by the latitude line at $\ph_0$ is $2\pi(1-\sin\ph_0)$, and the Gaussian curvature of the unit sphere is $1$, so the total geodesic curvature is $2\pi - 1\cdot2\pi(1-\sin\ph_0) = 2\pi\sin\ph_0$. Since the length of the latitude line is $2\pi\cos\ph_0$, the geodesic curvature at each point is
\begin{equation*}
    \frac{2\pi\sin\ph_0}{2\pi\cos\ph_0} = \tan\ph_0\tc
\end{equation*}
so $\th\prm(0) = \tan\ph_0$. Combining this fact with \cref{dduHR,dRdu}, we finally get
\begin{align*}
    &\,\pdpd{}{u}[H(\vn)\,R(u,v)]\Big|_{\vn=\vn_0} \\
    &= \frac{1}{\cos \ph_0}\,\pdpd{H}{\lm}(\lm_0,\ph_0) + \tan(\ph_0)\,H(\lm_0,\ph_0)\begin{bmatrix}
        0 & 1 \\ -1 & 0
    \end{bmatrix}\tp
\end{align*}
This completes the calculation of the derivatives of $\tl{K}$, which completes the calculation of $\del\Cost(\Verts)$.

Unlike for the sphere cartogram, it does not make sense to manually rotate the sphere after optimizing, because that would reintroduce distortion that we worked hard to optimize away. This means that, if we want the interruption to remain in a specific place, we must enforce this by modifying the cost function. For our cartograms, we place the interruption at $169^\circ$ west, through the Bering Strait. We rotate the borders so that the Cartesian coordinates on this line satisfy $y = 0$, which makes it simple to compute the deviation of points from this line.

We add a term to the definition of $\Dist(\Verts)$ that measures the deviation of points on the antimeridian from the interruption line. (Since the mesh is based on an octahedron, there is a line of mesh vertices that falls on the antimeridian.) We also add a term that fixes the North Pole in place. Let $M$ be the set of all vertices initially on the antimeridian (not including the poles), and let $\vp^\mathrm{north}$ be the vertex initially at the North Pole. We then modify the definition of the total distortion \cref{totaldist} to be
\begin{align*}
    \Dist(\Verts) = &\sum_{T \in \Tris} \im_T \bigrou{\wshp_T \dshp_T + \wscl_T \dscl_T} \\
    & + w^\mathrm{pole}\bigrou{(\vp^\mathrm{north}_x)^2 + (\vp^\mathrm{north}_y)^2} + \sum_{\vv \in M} w^\mathrm{antimer}_\vv \vv_y^2\tc
\end{align*}
where $w^\mathrm{pole}$ and $(w^\mathrm{antimer}_\vv)_{\vv\in M}$ are nonnegative weights. We set $w^\mathrm{pole}$ to be very high so that the North Pole is fixed, which helps to maintain the shapes of regions near the pole. We set $w^\mathrm{antimer}_\vv$ highest for vertices in the north, to ensure the interruption passes through the Bering Strait. We set $w^\mathrm{antimer}_\vv$ lower for vertices in the mid latitudes, and set $w^\mathrm{antimer}_\vv = 0$ around Antarctica.

To create the Liquid Earth projection (\cref{hybrid1}), we simply ignore the cartographic error, minimizing $\Dist(\Verts)$ instead of $\Cost(\Verts)$. We set the intended scale $s_T$ of each triangle to $1$. We weight triangles than contain land $100$ times more than water triangles, otherwise weighting different triangles equally. To produce a nearly equal-area map with minimal shape distortion, we minimize multiple times, each time weighting scale distortion higher. We start with scale distortion weighted $0.1$ times more than shape distortion and end with scale distortion weighted $100$ times more than shape distortion. Prioritizing shape distortion in the beginning is necessary to find a good local minimum. We do most of the optimization using an octahedral mesh of $18432$ triangles ($8 \cdot 48^2$), then subdivide in particularly distortion-prone places, then finish the optimization. To improve the look of the graticule, we fix the North Pole exactly in place, similar to how we fix the mesh vertices to the sphere in our sphere and hybrid cartograms.

\section{Algorithm performance}\label{performance}
In this section, we briefly lay out the runtime and accuracy of our cartogram algorithm.

For the tables below, we calculate relative cartographic error for each region $R$ as $(\mu_R - p_R)/p_R$, as described in previous sections. For the sphere and hybrid cartograms, the computation of the transformed area $\mu_R$ is based on the transformed mesh without considering the projection to the sphere. This projection step introduces a small amount of relative error that the tables do not capture.

To estimate this error, recall that the area of each triangle is computed after projecting the vertices to the tangent plane on the triangle's midpoint $\vn$. Consider an infinitesimal neighborhood of a point $\vv^\mathrm{tp}$ on this triangle in the tangent plane. Using similar triangles, it can be seen that the projection from the plane triangle to the spherical triangle scales this neighborhood by a factor of $\Ver{\vv^\mathrm{tp}} = \sqrt{1 + \Ver{\vv^\mathrm{tp} - \vn}^2}$. The relative area error of the projection on this neighborhood is then $\sqrt{1 + \Ver{\vv^\mathrm{tp} - \vn}^2} - 1 \approx \Ver{\vv^\mathrm{tp} - \vn}^2 / 2$. This means the maximum error incurred on this triangle is on the order of the area of the triangle. Since the mesh was subdivided so that the transformed triangle areas are less than $4\pi/2048$, and since the squared radius of an equilateral triangle is $4/3^{3/2}$ times its area, the maximum (relative) error incurred is less than $0.003$ in most cases.

The runtimes below do not include the time taken to compute the area portion values $\por{R,T}$. Depending on the resolution of the region border data, this can take several hours in Python. However, since the code for this is not vectorized, it could be greatly sped up by using a compiled programming language. Creating the meshes takes a negligible amount of time.

As described in \cref{overview}, we do the optimization in stages, minimizing $\Cost(\Verts) = \Werr\,\Err(\Verts) + \Wdist\,\Dist(\Verts)$ multiple times, setting $\Wdist$ to be smaller each time. At each stage, finding a local minimum of $\Cost(\Verts)$ requires thousands of search steps, stopping when $\del\Cost(\Verts) = 0$ is achieved. A key performance improvement is possible by not requiring a true local minimum to be found, instead stopping when the size of the gradient is below a certain threshold. At each stage, we choose a threshold value $\gm > 0$ and stop the optimization when $\Ver{\del\Cost(\Verts)}_\infty < \gm$. (For the sphere and hybrid cartograms, we remove the off-sphere component of $\del\Cost(\Verts)$ before doing this check.) For the first stage, we choose $\Werr = 1$, $\Wdist = 0.1$, and $\gm = 0.01$. Going from one stage to the next, we multiply $\Wdist$ and $\gm$ by $0.1$ while keeping $\Werr$ fixed.

Our implementation uses NumPy; the runtimes below were recorded for an AMD Ryzen 5 5600 CPU. We may sacrifice accuracy for speed by stopping after a smaller number of stages. The last few stages are unnecessary for the sphere and hybrid cartograms because of the aforementioned error caused by projection.
\begin{center}
\textbf{Plane cartogram}
\begin{tabular}{c|c c c c}
    Stage & Steps & Total time & Median rel.\ error & Max rel.\ error\\
    \hline
    1 & 76375 & 1h 4m 17s & 0.0425 & 4.19 \\
    2 & 2626 & 1h 6m 27s & 0.00472 & 3.91 \\
    3 & 4345 & 1h 9m 58s & 0.000502 & 2.59 \\
    4 & 304 & 1h 10m 13s & $4.56\cdot10^{-5}$ & 1.06 \\
    5 & 6331 & 1h 15m 20s & $4.81\cdot10^{-6}$ & 0.256 \\
    6 & 12212 & 1h 23m 40s & $4.89\cdot10^{-7}$ & 0.0355 \\
    7 & 248 & 1h 23m 50s & $6.60\cdot10^{-8}$ & 0.00377 \\
    8 & 4209 & 1h 26m 37s & $6.14\cdot10^{-9}$ & 0.000378 \\
    9 & 232 & 1h 26m 46s & $6.40\cdot10^{-10}$ & $3.78\cdot10^{-5}$ \\
    10 & 309 & 1h 26m 59s & $4.71\cdot10^{-11}$ & $3.78\cdot10^{-6}$ \\
\end{tabular}
\end{center}

\begin{center}
\textbf{Sphere cartogram}
\begin{tabular}{c|c c c c}
    Stage & Steps & Total time & Median rel.\ error & Max rel.\ error\\
    \hline
    1 & 48362 & 1h 19m 38s & 0.0410 & 3.79 \\
    2 & 4369 & 1h 26m 39s & 0.00446 & 3.57 \\
    3 & 526 & 1h 27m 27s & 0.000472 & 2.53 \\
    4 & 1391 & 1h 29m 43s & $5.05\cdot10^{-5}$ & 1.04 \\
    5 & 207 & 1h 30m 0s & $4.95\cdot10^{-6}$ & 0.262 \\
    6 & 2270 & 1h 32m 52s & $5.53\cdot10^{-7}$ & 0.0367 \\
    7 & 361 & 1h 33m 19s & $6.62\cdot10^{-8}$ & 0.00389 \\
    8 & 2431 & 1h 36m 19s & $6.56\cdot10^{-9}$ & 0.000391 \\
    9 & 202 & 1h 36m 34s & $5.54\cdot10^{-10}$ & $3.92\cdot10^{-5}$ \\
    10 & 108 & 1h 36m 43s & $4.64\cdot10^{-11}$ & $3.92\cdot10^{-6}$ \\
\end{tabular}
\end{center}
\begin{center}
\textbf{Hybrid cartogram}
\begin{tabular}{c|c c c c}
    Stage & Steps & Total time & Median rel.\ error & Max rel.\ error\\
    \hline
    1 & 69268 & 2h 9m 51s & 0.0518 & 4.64 \\
    2 & 9289 & 2h 27m 12s & 0.00578 & 4.28 \\
    3 & 5105 & 2h 37m 1s & 0.000657 & 2.95 \\
    4 & 4649 & 2h 45m 41s & $5.88\cdot10^{-5}$ & 1.20 \\
    5 & 9197 & 3h 2m 42s & $6.05\cdot10^{-6}$ & 0.303 \\
    6 & 35513 & 3h 58m 5s & $6.60\cdot10^{-7}$ & 0.0442 \\
    7 & 6575 & 4h 8m 8s & $7.03\cdot10^{-8}$ & 0.00471 \\
    8 & 1476 & 4h 10m 22s & $7.54\cdot10^{-9}$ & 0.000474 \\
    9 & 902 & 4h 11m 44s & $7.60\cdot10^{-10}$ & $4.74\cdot10^{-5}$ \\
    10 & 772 & 4h 12m 59s & $6.44\cdot10^{-11}$ & $4.74\cdot10^{-6}$ \\
\end{tabular}
\end{center}

\section{Final Remarks}\label{remarks}
The algorithm we present in this paper gives a significant improvement in shape preservation over existing methods. It achieves very low cartographic error in the sphere and hybrid cases and arbitrarily low error in the plane case, barring floating-point issues. Potential areas of future research include improving the algorithm's runtime and testing whether the resulting cartograms are rated favorably by map readers.

It is unclear how much mesh subdivision needs to be done to guarantee that an accurate cartogram is possible. For example, if two regions lie entirely within a single triangle, it is clearly impossible to make both regions' areas match their populations. Finding a sufficient condition for area accuracy to be possible could be of independent mathematical interest. It is also unclear why the weight on scale distortion has a strong effect on runtime. Resolving these questions is a potential area of future research.

The on-sphere techniques presented here could be applied to other cartogram algorithms. Though Li and Aryana \cite{sphere} adapted the diffusion method to the sphere, the same has not been done for rubber-sheet cartograms. Such on-sphere adaptations of cartogram methods could also be made to take into account the projection to the plane, similar to our hybrid cartograms.


\begin{thebibliography}{10}
    \bibitem{line} Armijo, L.\ (1966). Minimization of functions having Lipschitz continuous first partial derivatives. \textit{Pacific Journal of Mathematics}, 16(1), 1--3. \url{https://doi.org/10.2140/pjm.1966.16.1}
    \bibitem{dcn} Dougenik, J.\ A., Chrisman, N.\ R., Niemeyer, D.\ R.\ (1985). An algorithm to construct continuous area cartograms. \textit{The Professional Geographer}. \url{https://doi.org/10.1111/j.0033-0124.1985.00075.x}
    \bibitem{gn} Gastner, M., Newman, M.\ E.\ J.\ (2004). Diffusion-based method for producing density-equalizing maps. \textit{Proceedings of the National Academy of Sciences}. \url{https://doi.org/10.1073/pnas.0400280101}
    \bibitem{quant} Jawaherul Alam, M., Kobourov, S.\ G., Veeramoni, S.\ (2015). Quantitative Measures for Cartogram Generation Techniques. \textit{Computer Graphics Forum}. \url{https://doi.org/10.1111/cgf.12647}
    \bibitem{man} Kronenfeld, B.\ J.\ (2017). Manual construction of continuous cartograms through mesh transformation. \textit{Cartography and Geographic Information Science}, 45(1), 76--94. \url{https://doi.org/10.1080/15230406.2016.1270775}
    \bibitem{kun} Kunimune, J.\ H.\ (2020). Minimum-error world map projections defined by polydimensional meshes. \textit{International Journal of Cartography}. \url{https://doi.org/10.1080/23729333.2020.1824174}
    \bibitem{elastic} Kunimune, J.\ H.\ (2023). Introducing the Elastic projections. Web. \url{https://kunimune.blog/2023/12/29/introducing-the-elastic-projections/}
    \bibitem{sphere} Li, Z., Aryana, S.\ (2017). Diffusion-based cartogram on spheres. \textit{Cartography and Geographic Information Science}, 45:5, 464--475. \url{https://doi.org/10.1080/15230406.2017.1408033}
    \bibitem{gatt} Loncaric, M.\ (2024). Map Projections 2: Solving Numerically. Web. \url{https://graphallthethings.com/posts/map-projections-2}
    \bibitem{l-bfgs} Nocedal, J.\ (1980). Updating quasi-Newton matrices with limited storage. \textit{Mathematics of Computation}, 35(151), 773--782. \url{https://doi.org/10.1090/S0025-5718-1980-0572855-7}
    \bibitem{cart} Nusrat, S., Kobourov, S.\ (2016). The state of the art in cartograms. \textit{Computer Graphics Forum}. \url{https://doi.org/10.1111/cgf.12932}
    \bibitem{liquid} Sargent, R.\ C.\ (2024). Introducing the Liquid Earth projection. Web. \url{https://rsargentmath.github.io/posts/liquid_earth/}
    \bibitem{odcn-a} Sun, S.\ (2013). A fast, free-form rubber-sheet algorithm for contiguous area cartograms. \textit{International Journal of Geographic Information Science}, 27 (3): 567--93. \url{https://doi.org/10.1080/13658816.2012.709247}
    \bibitem{odcn-b} Sun, S.\ (2013). An optimized rubber-sheet algorithm for continuous area cartograms. \textit{The Professional Geographer}. \url{https://doi.org/10.1080/00330124.2011.639613}
    \bibitem{odcn-c} Sun, S.\ (2015). Carto3F program: A fast, free-form algorithm implementation for area
cartograms. Web. \url{http://www.sunsp.net/download.html}
\end{thebibliography}
\end{document}